\documentclass[showpacs,aps]{revtex4}
\usepackage{graphicx} 
\usepackage{bm}
\newcommand{\mq}{\mu_q}
\newcommand{\tc}{T_\chi}
\newcommand{\uu}{\langle \bar{u} u \rangle}

\newcommand{\ber}{\begin{eqnarray}}
\newcommand{\eer}{\end{eqnarray}}
\newcommand{\beq}{\begin{equation}}
\newcommand{\eeq}{\end{equation}}
\newcommand{\sgh}{\sigma_M(\omega, \vec{k})}
\begin{document}
\begin{flushright}
TIFR/TH/08-57 \\ 
CU-PHYSICS/1-2009
\end{flushright}
\title{Mesonic Excitations of QGP: Study with an Effective Model}
\author{Paramita Deb}
\email{paramita.deb83@gmail.com}
\author {Abhijit Bhattacharyya}
\email{abphy@caluniv.ac.in}
\affiliation{Department of Physics, University of Calcutta,
92, A. P. C. Road, Kolkata - 700009, INDIA}
\author{Saumen Datta}
\email{saumen@theory.tifr.res.in}
\affiliation{Department of Theoretical Physics, Tata Institute
of Fundamental Research, Homi Bhabha Road, Mumbai - 400005, INDIA}
\author{Sanjay K. Ghosh}
\email{sanjay@bosemain.boseinst.ac.in}
\affiliation{Centre for Astroparticle Physics \&
Space Science and Department of Physics, Bose Institute,
93/1, A. P. C Road, Kolkata - 700009, INDIA}

\vskip 0.3in

\begin{abstract}
We study the correlations between quark-antiquark pairs in different quantum
number channels in a deconfined plasma by using an effective model of QCD.
Using the three flavour PNJL model, the finite temperature spectral functions 
for different mesonic states are studied at zero and nonzero quark chemical 
potentials. It is found that in the $\eta$ channel resonance 
structures 
survive above the chiral transition temperature $\tc$, while the 
kaonic states seem to get washed off just above $\tc$.
The sensitivity of the structures to the 
anomaly term are carefully investigated.
\end{abstract}

\maketitle

{\section {INTRODUCTION}}
\label{sec.intro}

At high temperatures and densities, strongly interacting matter is expected 
to undergo a transition to new phases where colour degrees of freedom are 
deconfined. Such conditions of high temperatures 
and densities can be created in the laboratory by the collision of heavy 
ions at large energies. A large amount of data have already been obtained
from RHIC. In near future results from LHC, for low density and high temperature phase,
will be available as well. On the other hand, 
the FAIR experiment at GSI will provide us information about extremely 
dense matter at low temperature. It is thus extremely relevant to study 
the properties of strongly interacting matter at high temperature and density.

A first principle study of hot and dense strongly interacting matter 
starting from QCD is not easy because the physics is nonperturbative in
the temperature and density range of interest. At finite temperatures,
numerical studies on lattice provide the most reliable results for the 
physics of the deconfined phase. The physics of excitations in real time, 
however, is not directly accessible from lattice and one requires an
analytical continuation from Euclidean time. Studies at finite baryon
number density are prohibitively difficult on the lattice, because of
the well-known sign problem. Some progress has been made in recent
years in extracting results for small densities 
\cite{fodor,philipsen}. As it stands now, the lattice 
predicts that the deconfinement and the chiral symmetry restoration 
happen roughly simultaneously \cite{tc} via a smooth crossover \cite{aoki} 
at zero quark chemical potential,
and the two transitions roughly coincide at about $\tc \sim 180-200$ MeV
 \cite{tc}. There is an indication of a critical point at a small
 chemical potential $\mu_B \approx$ 300 - 500 MeV, with a first order transition
 for larger $\mu_B$ \cite{fodor}. There have also been suggestions that the 
first order line may not exist \cite{philipsen}. 

In view of the limitations in studying finite density and real time 
problems, it may be useful to get an idea of the features 
of the new phases of matter by studying QCD inspired models. Depending on
the physics of interest, one can attempt to study a model which has features 
relevant to the physics issue. In particular, for issues of the transition of
the hadronic matter to a chiral symmetry restored, deconfined phase the 
Polyakov loop extended Nambu -- Jona-Lasinio (PNJL) model has been used.

In the Nambu -- Jona-Lasinio (NJL) model \cite{klev,hatsuda}, as it is
used currently, the interactions between quarks are taken into account
by suitable four-quark terms which respect the chiral symmetry of the
QCD lagrangian. Since the gluons are integrated out, with their only
effect being the four-quark interaction terms, there is no confinement
per se in this model. The chiral symmetry is, however, spontaneously
broken by ${\bar q} q$ condensation: $U_L(N_f) \times U_R(N_f) \to U_V(N_f)$.  
As a result, the quarks pick up a constituent mass. Furthermore, 
the pseudoscalar mesons $\pi, K$ and $\eta$ become the Goldstone modes. 
It is also easy to incorporate the
$U_A(1)$ anomaly by introducing a suitable determinant term \cite{hatsuda}.
The NJL model has been widely used to study the chiral phase
transition in QCD, and also the nature of the excitations at
temperatures slightly above the transition temperatures.
Based on such studies, it was suggested sometime back that nontrivial 
strong correlations between $q \bar{q}$ pairs persist in the deconfined 
phase at moderately high temperatures \cite{hatsuda}.

Due to the lack of confinement, though, the NJL model does miss out on
some important aspects of the QCD thermal transition. In particular,
the chiral transition in QCD is also of deconfining nature, as the
Polyakov loop, which is the confinement-deconfinement order parameter, 
shows a rapid change. The Polyakov loop extended NJL (PNJL) model
\cite{fuku} attempts to capture this feature of the QCD transition.
In this model, the gluon dynamics 
is described by a background temporal gluon field which is coupled 
to the quarks by the covariant derivative. The value of 
the background field is determined by minimizing the corresponding 
potential $U(\Phi)$
which depends on the traced polyakov loop $\Phi$. 

The PNJL model has been extensively used to look at the phase structure at 
high temperature and density. Ratti {\it et.al.} \cite{ratti} have studied 
the two flavour 
version of this  model in certain amount of detail. They have looked at the 
crossover transition around 200 MeV and also the quark number densities 
at different baryonic chemical potentials. Their results match nicely with 
the lattice, especially the fact that the chiral restoration 
and the deconfinement transition take place almost simultaneously. 
The speed of sound and the diagonal susceptibilities, calculated with the PNJL
model, were also found to be in good agreement with lattice results 
\cite{ghosh}. On the other hand the off-diagonal susceptibilities in two 
flavour PNJL model differ from the lattice results \cite{swa}.
Recently Fukushima has studied the phase diagram of the
three flavour quark matter within the framework of the PNJL model
\cite{fukushima2}.

Since the PNJL model successfully reproduces a large number 
of quantities calculable directly from QCD one may be interested to 
use this model to study quantities of experimental interest which
cannot be directly studied from lattice. In particular, a knowledge of
the nature of excitations of the plasma is phenomenologically very
important. The complete in-medium behavior of the mesonic excitations 
require a study of the spectral functions in these channels. As we
discussed earlier, it is not easy to calculate them on lattice. A
detailed study of such spectral functions in the NJL model has been 
carried out sometime back \cite{hatsuda}. It is clearly of interest 
to find out how the introduction of the Polyakov loop changes the 
nature of the $q \bar{q}$ correlations. Hansen {\it et.al.} 
\cite{hansen} have studied the pion and sigma correlation functions in
the 2-flavour PNJL model. Costa {\it et.al.} \cite{costa1} have calculated the 
masses of the pseudoscalar mesons at finite temperature and zero 
density. It is, of course, interesting and
phenomenologically important to study a larger set of mesonic
correlations for the realistic case of 2+1 flavour PNJL model and also 
at finite density.  

Here we aim to carry out such a study. We look at the spectral
functions and pole masses of the pseudoscalar channels with different
flavour contents. We also carefully 
investigate the effect of the anomaly term in the spectral functions. 
The spectral functions have also been studied in presence of a small 
chemical potential. 

This paper is organised as follows: in the next section we briefly describe 
the 2+1 flavour PNJL model, as used by us. In Sec. \ref{sec.phase} the basic 
in-medium calculations and the chiral transition have
been discussed. The main results of the paper, about mesonic
excitations, are in Sec. \ref{sec.meson}. A summary of our results and
conclusions are available in the last section.

\vskip 0.3in    
\section {Three Flavour PNJL Model}
\label{sec.model}

In this section, we briefly describe the PNJL model, and specify the 
parameters of the model that we have used. More details can be found in 
the literature \cite{fuku,ratti,gatto}.

In the NJL model the gluon dynamics is reduced to the chiral point 
couplings between quarks. The PNJL model also introduces the temporal
gauge field, since the Polyakov loop,
\begin{equation}
   \Phi = {1 \over V} \int d^3x \ \ {1 \over N_c} \ {\rm Tr} \ {\cal P} 
{\rm exp} [i {\int_0}^\beta d\tau A_4{({\bar x},\tau)}]
\label{eq.polloop}
\end{equation}
is the order parameter for the confinement-deconfinement transition.
 Here $A_4=iA_0$ is the temporal component of Euclidean gauge field, 
$\beta=1/T$, and $\cal P$ denotes path ordering.
   Our effective
 $SU(3)_f$ lagrangian is \cite {gatto}
\begin {eqnarray}
   {\cal L} &=& {\sum_{f=u,d,s}}{\bar\psi_f}\gamma_\mu iD^\mu
             {\psi_f}-\sum_f m_{f}{\bar\psi_f}{\psi_f}
              +\sum_f \mu \gamma_0{\bar \psi_f}{\psi_f}
       +{{g_S}\over 2} {\sum_{a=0,\ldots,8}}[({\bar\psi} \lambda^a {\psi})^2+
            ({\bar\psi} i\gamma_5\lambda^a {\psi})^2]\nonumber\\
       &{}& \qquad \qquad -{g_D} [det{\bar\psi_f}(1+\gamma_5){\psi_{f^\prime}}
       +det{\bar\psi_f}(1-\gamma_5){\psi_{f^\prime}}]
           -{\cal U}(\Phi[A],\bar \Phi[A],T)\nonumber\\
  &=& {\cal L}_0+{\cal L_{M}}+{\cal L}_\mu+{\cal L}_s+{\cal L_{KMT}}-{\cal U}
\label{lag}
\end {eqnarray}
In the above Lagrangian ${\cal L}_0$ is the kinetic term with gauge field 
interactions, $D^\mu=\partial^\mu-iA_4 \delta_{\mu 4}$. The gauge coupling 
is 
absorbed in the definition of $A^\mu$. ${\cal L}_{M}$ is the 
mass term which breaks the chiral symmetry explicitly. The mass of a particular 
flavour is denoted by $m_f$ and the corresponding field is $\psi_f$. The
light quark mass is considered to be 
$\sim 5 $MeV and the strange quark mass is considered to be $\sim 140$ MeV. 
The term ${\cal L}_s$ is responsible for the four-fermi interaction 
among the quarks with coupling $g_s$. Here we take this coupling to be 
positive. The next term ${\cal L}_{KMT}$ which is a six-fermi interaction is 
 invariant under ${SU(3)}_L\times {SU(3)}_R$ but breaks $U(1)_A$ symmetry. 
This term represents the axial anomaly of QCD \cite{hatsuda}. Here 'det' stands
for the determinant with respect to the flavour indices, and the anomalous 
coupling is represented as $g_D$. This anomaly term is responsible for the 
flavour mixing of $\eta_0$ and $\eta_8$ mesons in the pseudoscalar channel,
giving rise to the $\eta$ and $\eta^\prime$ mesons. The potential 
$U(\Phi)$ is expressed in 
terms of the Polyakov loop $\Phi$ and its conjugate, $\bar \Phi$, 
as \cite{ratti}.
\begin{equation}
  {{\cal U}(\Phi, \bar \Phi, T) \over {T^4}}=-{{b_2}(T) \over 2}
                 {\bar \Phi}\Phi-{b_3 \over 6}(\Phi^3 + \bar \Phi^3)
                 +{b_4 \over 4}{(\bar\Phi \Phi)}^2
\label{eq.polpot}
\end{equation}
where
\begin {equation}
     {b_2}(T)=a_0+{a_1}({{T_0}\over T})+{a_2}({{T_0}\over T})^2+
              {a_3}({{T_0}\over T})^3,
\end {equation}
$b_3$ and $b_4$ being constants and $T_0$, which is a parameter here, is 
chosen to be $190$ MeV.

Both $\Phi$ and ${\bar \Phi}$ are treated as classical field variables. 
When the quark number density is zero one has $\Phi = {\bar \Phi}$. This 
quantity can be considered as the order parameter for the phase 
transition. 
Furthermore $U[\Phi,{\bar \Phi},T]$ has a $Z(3)$ center symmetry 
which encompasses the phase transition in QCD. At low temperature 
$U$ has a single minimum at $\Phi = 0$. At high temperature $U$ has 
three degenerate minima at $\Phi=1,e^{\pm 2i\pi/3}$. 

To study the chiral transition, we study the system of eqn.(\ref{lag}) 
 in the Mean Field Approximation (MFA)
 to get the field equations for $\Phi$, $\bar \Phi$, $\sigma$.
Due to the breaking of ${SU(3)}_L \times {SU(3)_R}$ symmetry to 
$SU(3)_V$ the quark condensate acquires a non-zero vacuum expectation value 
given by 
\begin {equation}
 <{\bar \psi_f}{\psi_f}>= -i{N_c}{{{\cal L}t}_{y\rightarrow x^+}}({\rm Tr}
\  {S_f}(x-y))
\end {equation}
where trace is over colour and spin states. 
Also, eight goldstone bosons appear for $N_f=3$ model. 
The self-consistent
gap equation for the constituent masses are
 \begin {eqnarray}
  M_u &=&m_u-4g_S \sigma_u+2g_D \sigma_d\sigma_s \nonumber \\
  M_d &=&m_d-4g_S \sigma_d+2g_D \sigma_s\sigma_u \nonumber \\
  M_s &=&m_s-4g_S \sigma_s+2g_D \sigma_u\sigma_d
\label{eq.mass}
\end {eqnarray}
Here $\sigma_f=<{\bar \psi_f} \psi_f>$ denotes the chiral condensate of
a quark with flavour $f,$ where $f=u, d, s$. 
The expression for $\sigma_f$ at $T=0$ and $\mu=0$ can be written as
 \cite{gatto}
\begin {equation}
 \sigma_f=-{3{M_f}\over {\pi}^2} {{\int_0}^\Lambda}{p^2\over
           \sqrt {p^2+{M_f}^2}}dp,
\end {equation}
$\Lambda$ being the three momentum cut off.
 
The parameters of the NJL part of the Lagrangian are fixed at $T=\mu=0$. 
In the present work we 
have used the parameter set obtained in ref. \cite{hatsuda},
\begin {eqnarray}
\Lambda=631.4 {\rm MeV}, {\hspace{0.05in}}m _u=m_d=5.5 {\rm MeV}, {\hspace{0.05in}} 
m_s=135.7 {\rm MeV}, {\hspace{0.05in}} g_S {\Lambda}^2= 3.67, 
{\hspace{0.05in}} g_D {\Lambda}^5= 9.29 \nonumber
\end {eqnarray}
This parameter set reproduces the physical values of the pion mass, 
$m_{\pi}=138 $ MeV, pion decay constant, $f_{\pi}=93$ MeV,
and masses of the $K$ and ${\eta}^\prime$ mesons, 
$m_K=495.7$ MeV and $m_{{\eta}^\prime}=957.5$ MeV.
 The parameters of $U(\Phi)$ are\cite{ratti,gatto},
\begin {eqnarray}
a_0=6.75,{\hspace{0.05in}}a_1=-1.95,{\hspace{0.05in}}a_2=2.625,
{\hspace{0.05in}}a_3=-7.44,{\hspace{0.05in}}b_3=0.75,{\hspace{0.05in}} 
b_4=7.5, T_0 = 190 {\rm MeV} \nonumber
\end {eqnarray}
so as to reproduce the lattice measurements of the Polyakov loop 
expectation value.

\vskip 0.3in

\section{Chiral transition in PNJL model}
\label{sec.phase}
In the mean field approximation, the thermodynamic potential of the 
PNJL model can be written as \cite{gatto}
\begin {equation} 
  \Omega (\Phi,\bar \Phi,M,T,\mu)= {\cal U}[\Phi,\bar \Phi,T]+2{g_S}
           {\sum_{f=u,d,s}}{{\sigma_f}^2}-4{g_D}{\sigma_u}{\sigma_d}{\sigma_s}
                -T{\sum_n}\int {{d^3p}\over{(2{\pi})^3}}
                {\rm Tr} \ln{{ S^{-1}}(i{\omega_n},\bar p)\over T}
\end {equation}
 where $S^{-1}$ is the inverse quark propagator, 
 \begin {equation}
   {S^{-1}}(p_0,\vec{p})=\gamma_0(p^0+\mu-i{A_4})-\vec{\gamma}\cdot{\vec p}-M
\end {equation}
and $\omega_n=\pi T(2n+1)$ are the Matsubara frequencies of fermions. 
Using the identity ${\rm Tr}\ln\left(X\right)=\ln \det\left(X\right)$, 
and performing the colour trace, we get
\begin {eqnarray}
 \Omega &=& {\cal U}[\Phi,\bar \Phi,T]+2{g_S}{\sum_{f=u,d,s}}
            {{\sigma_f}^2}-4{g_D}{\sigma_u}
          {\sigma_d}{\sigma_s}-6{\sum_f}{\int_{0}^{\Lambda}} 
     {{d^3p}\over{(2\pi)}^3} E_{pf}\Theta {(\Lambda-{ |\vec p|})}\nonumber \\
       &{}& -2{\sum_f}T{\int_0^\infty}{{d^3p}\over{(2\pi)}^3}
           \left(
            \ln\left[1+3(\Phi+{\bar \Phi}e^{-{(E_{pf}-\mu)\over T}})
      e^{-{(E_{pf}-\mu)\over T}}+e^{-3{(E_{pf}-\mu)\over T}}\right] \right.
\nonumber\\
    &{}& \qquad \qquad \qquad + \left. \ln\left[1+3({\bar \Phi}+{ \Phi}e^{-{(E_{pf}+\mu)\over T}})
            e^{-{(E_{pf}+\mu)\over T}}+e^{-3{(E_{pf}+\mu)\over
		T}}\right] \right)
\label{eq.mfpot}
\end {eqnarray}
where $E_{pf}=\sqrt {p^2+M^2_f}$ is the single quasiparticle energy. 
In the above integrals, the vacuum integral
has a cutoff $\Lambda$ whereas the medium dependent integrals
have been extended to infinity. 

\vskip 0.1in

\begin{figure}[htb]
\vskip 0.2in
\centerline{\includegraphics[width=6in]{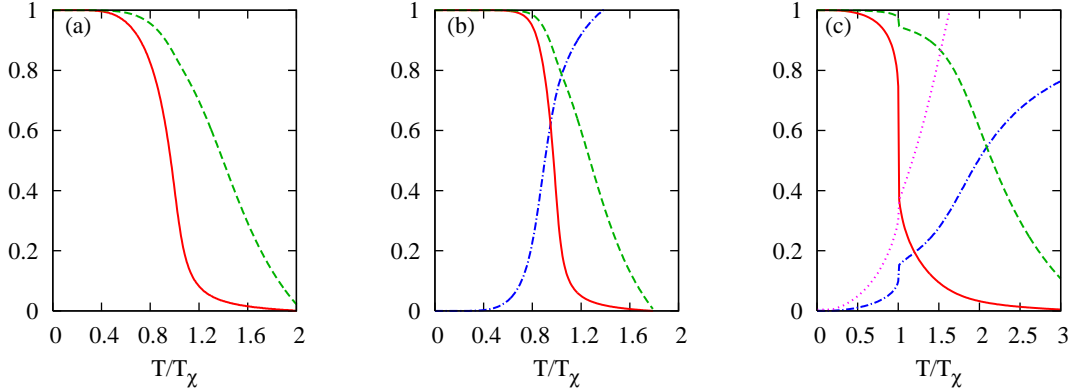}}
\caption{(Color online) (a) $\langle \bar{u} u \rangle (T) / 
\langle \bar{u} u \rangle (T=0)$ (red, solid) and 
$\langle \bar{s} s \rangle (T) / \langle \bar{s} s \rangle (T=0)$ 
(green, dashed), as a function of $T/\tc$  at $\mu_q = 0$ 
for the NJL model. (b) The same for the PNJL 
model; also shown is the Polyakov loop expectation value, $\Phi$ (blue, 
dash-dotted). (c) Same as (b) but for 
$\mu_q = 320 $MeV.  Here $\Phi$ (blue, dash-dotted) and ${\bar \Phi}$ 
(pink, dotted) are different.}
\label{fig.uu} \end{figure}

To look at thermal behavior of different observables, we need to study 
the variation with temperature of the minimum of the thermodynamic
potential. In fig. \ref{fig.uu} we show the thermal behavior of the 
chiral condensates for both the light and the strange quarks, in both
NJL and PNJL models. The NJL results can be obtained from
eq. (\ref{eq.mfpot}) by setting $U[\Phi,\bar \Phi, T]=0$, $\Phi=1$ and 
$\bar \Phi=1$. Fig. \ref{fig.uu} shows a rapid drop in $\uu$ in both
the models in the temperature range 150-200 MeV, without any
singularity, indicating a chiral symmetry restoring crossover. This is
consistent with what lattice studies find for QCD \cite{aoki}. From
the points of inflexion of $\uu (T)$  one estimates $\tc \sim$ 170 MeV 
for NJL and 190 MeV for PNJL models. Figure \ref{fig.uu} also shows
the behavior of the Polyakov loop, $\Phi$ \cite{note3}. The figure shows that
deconfinement temperature, $T_d \sim \tc$, as expected from lattice
studies of QCD \cite{tc}. 

At large quark densities, the chiral symmetry restoring transition in
PNJL model becomes first order. 
Figure \ref{fig.uu} also shows the behavior of the chiral condensate
and the Polyakov loop in the presence of a chemical potential $\mq$ =
320 MeV, which is just above the critical point for this model. 
The transition temperature at this $\mq$ is $\sim$ 85 MeV. Note the sharp fall
of $\uu$, indicating a first order transition. All the quantities shown in 
the figure show a jump at the transition point, consistent with a 
discontinuous transition, though the drop in $\langle \bar{s} s \rangle$ 
and the rise in $\Phi$ are only a few \%, and both these quantities have 
rather complicated temperature dependence above the transition point.  
In the presence of a
chemical potential, there is no charge conjugation symmetry so 
$\Phi \ne \bar{\Phi} $, as seen in
fig. \ref{fig.uu}. 

One problem of the mean-field study of PNJL model is the absence of color neutrality
at finite density \cite{abuki,abuki1}. This arises because of the fact that 
the Polyakov loop, which couples to the dynamical quarks, serves as an 
external coloured field, leading effectively to separate chemical potentials for 
quarks of different color. As a remedy to this problem, it has been suggested to 
introduce a colour chemical potential which will enforce color neutrality\cite{abuki}. 
While this affects the individual number densities of the quarks of different colors, 
the total number densities of quarks and antiquarks, $n_q$ and $n_{\bar q}$, 
do not change significantly \cite{abuki1}. We will discuss in 
Sec. \ref{sec.density} the implication of this for our study.

\begin{figure}[htb]
\vskip 0.2in
\centerline{\includegraphics[width=5.5in]{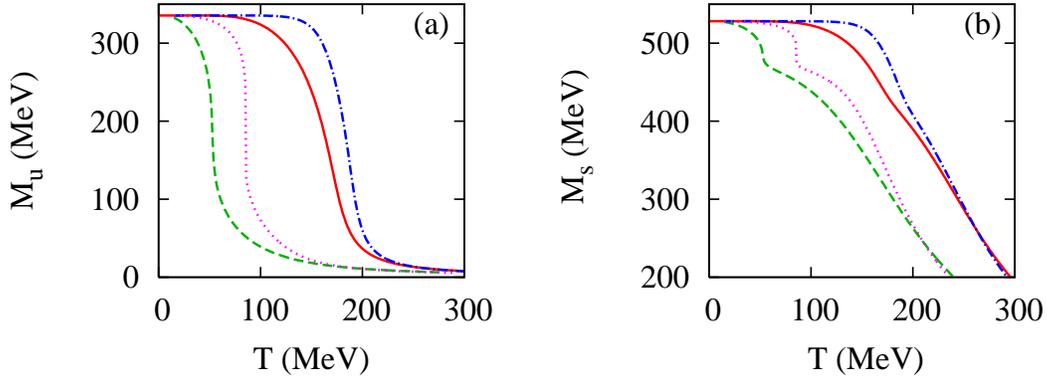}}
\caption{(Color online) (a) Variation of $M_u$ (a) with 
temperature, at $\mu=0$ and $320 $MeV for both NJL and PNJL model. Solid (red)
and dashed (green) lines are for NJL model with $\mu_q$ = 0 and 320 MeV, 
respectively, while dashed-dot (blue) lines and dotted (pink) lines 
show the corresponding figures for PNJL model. (b) Same as (a) but for 
$M_s$.} 
\label{mus}
\end{figure}

In fig.\ref{mus} we have shown the variation of $M_u$ and $M_s$
with temperature at different chemical potentials. Here $M_u$ and $M_s$
are the constituent quark masses, eq.(\ref{eq.mass}). At $\mq$ = 0,  
The change in the mass is smooth. But the change becomes sharper on
introducing a chemical potential. While the introduction of the Polyakov
loop changes the constituent quark masses considerably around the 
phase boundary, at very high temperature both the models give similar
results, since there $\Phi\rightarrow 1$. A similar variation of 
the strange quark mass is obtained in the two models.

\begin{figure}[htb]
\vskip 0.2in
\centerline{\includegraphics[width=8cm,height=6cm]{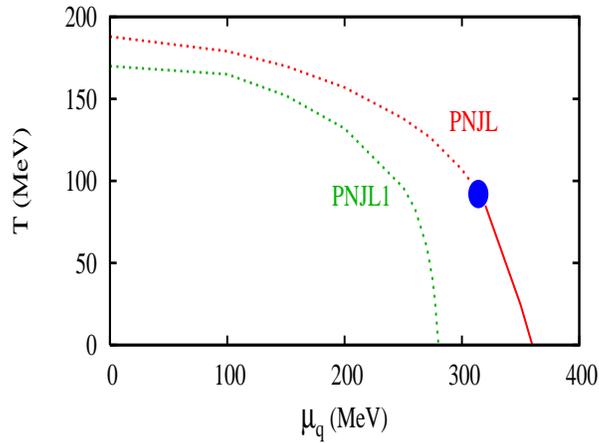}}
\caption{(Color online) Phase diagram in the PNJL model. The solid line at 
large 
$\mq$ ends in a critical point at $(\mu_c, T_c)$ = (314 MeV, 92 MeV). 
At smaller $\mq$ one gets a crossover, denoted by the dotted line. 
Putting in an explicit temperature dependence in the anomaly term
$g_D(T) = g_D(0) \ {\rm exp}(-T / \tc)^2$ (denoted PNJL1; see text)
washes off the first order line completely.}
\label{fig.phase}
\end{figure}

The phase diagram we obtain is shown in fig. \ref{fig.phase}. 
The critical point in our calculation was found to be
at $(\mu_c, T_c)$ = (314 MeV, 92 MeV). While lattice estimates of
the critical chemical potential $\mu_c$ vary considerably between
different groups, they tend to be much lower than the PNJL value,
$\leq$ 150 MeV. This probably indicates that the naive PNJL (or NJL) 
model may not be a good model for the QCD transition at large $\mq \sim$ 
300 MeV, since nuclear excitations become important in this regime. 

Another point to mention here is that the $U_A(1)$ anomaly term is introduced
explicitly in the NJL and PNJL models by a six fermion interaction
term. In QCD, of course, this term is introduced due to instantons, which
are topological objects in the gauge sector \cite{thooft}. 
It is not known how well the anomalous $U_A(1)$ symmetry is restored in 
the deconfined phase. In Ref. \cite{hatsuda} it was speculated that the 
$U_A(1)$ symmetry is restored well below $\tc$. While there is no 
evidence of that from lattice studies, it is an open question
whether the $U_A(1)$ symmetry is restored (or at least, the effect
of the anomaly term substantially reduced) at above $\tc$. 
If we include an explicit temperature dependence of the coefficient 
of the anomaly term, $g_D(T) = g_D(0) \ {\rm exp}(-T / \tc (mu))^2$,
where $\tc(\mu)$ is the transition temperature at quark chemical potential $\mu$, 
then the transition becomes considerably softer and the first order line 
completely disappears, while the transition temperature moves to a 
somewhat lower value (see fig. \ref{fig.phase}). We use PNJL1
to denote the model with such a temperature dependence of $g_D$ in 
eq. (\ref{lag}). One might be tempted to make a conclusion 
that the existence of critical point is strongly related to the strength 
of the instanton-induced anomaly term at the transition point \cite{note}. 

Of course, PNJL1 implies a substantial reduction of the anomaly term 
already before $\tc$. A more realistic temperature dependence of $g_D$ could be
a sharp drop around $\tc$. To investigate the effect of such a temperature 
dependence, we also consider eqn. (\ref{lag}) with $g_D(T) = g_D(0) \ 
(1-{\rm tanh}\ 5x)/2$, where $x=(T-\tc (\mu))/\tc (\mu)$. We call this model PNJL2 
\cite{note2}. PNJL2 has a phase diagram similar to PNJL {\it i.e.} the 
critical point has the same position at $(\mu_c, T_c)$ = (314 MeV, 92 MeV). 
However, as we will see later, the nature of the mesonic excitations around $\tc$ can be 
substantially different between the two. 

\vskip 0.3in
\section{Mesonic Excitations and $U_A(1)$ Anomaly}
\label{sec.meson}

In order to understand the properties of the medium beyond the bulk
thermodynamic properties, we need to look at the low-lying
excitations. The simplest gauge invariant excitations in QCD are the mesonic 
excitations. Masses and decay widths of the mesonic resonances 
are calculated from the correlations of $\bar{q} \Gamma q$-type operators
in QCD vacuum. In the medium,
we are interested in spectral changes due to interactions with the 
medium; in particular, we would be interested in possible resonance-like 
structures in the deconfined phase. We need therefore to look at the
spectral functions.

Here we study spectral functions of pseudoscalar and scalar 
mesonic states. They are of interest because of their close connection with 
the  chiral symmetry breaking and its restoration. The temperature
dependence of the spectral function has been studied
in the Nambu-Jona-Lasinio model in ref. \cite{hatsuda}. There it was found that 
narrow structures persist in the symmetry restored phase, at moderately 
high temperatures. Here we intend to see how the coupling with the Polyakov 
loop affects these results. Also we study the sensitivity of these structures
on the coefficients of the anomaly term. We also extend the studies to nonzero 
quark chemical potentials.

\subsection{Formalism}

The spectral function $\sgh$ for a given 
mesonic channel $M$ in a system at temperature T can be defined 
through the Fourier transform of the real time two point functions
$D^{>}$ and $D^{<}$ \cite{lebellac},
\ber
\sgh &=& \frac{1}{2 \pi} (D^{>}_M(k_0, \vec{k})-D^{>}_M(k_0, \vec{k})) 
\label{eq.defspect} \\
 D^{>(<)}_M(k_0, \vec{k}) &=& \int{d^4 x \over (2
\pi)^4} e^{i k.x} D^{>(<)}_M(x_0,\vec{x}) \nonumber \\
D^{>}_M(x_0,\vec{x}) &=& \langle
J_M(x_0, \vec{x}) J_M(0, \vec{0}) \rangle_c \label{eq.def2pt} \\
D^{<}_M(x_0,\vec{x}) &=& 
\langle J_M(0, \vec{0}) J_M(x_0,\vec{x}) \rangle_c \nonumber
\eer
Here $J_M$ is the suitable hadronic operator and the subscript $c$ denotes 
the connected part of the correlation function. 
Using the Kubo-Martin-Schwinger(KMS) condition \cite{lebellac}
\beq
D^{>}_M(x_0,\vec{x})=D^{<}(x_0+i/T,\vec{x})
\label{eq.kms}
\eeq
one can connect $\sgh$ to the retarded correlation function,
\ber
\sgh &=& 2 \pi \ {\rm Im} D^R_M(k_0, \vec{k}) \nonumber \\
D^R_M(k_0, \vec{k}) &=& \int{d^4 x \over (2
\pi)^4} e^{i k.x} \theta(x_0) \langle [J_M(x_0,\vec{x}),J_M(0,\vec{0}]
\rangle_c
\label{eq.retard} \eer
Inserting a complete set of
states in eq. (\ref{eq.defspect}) and using eq. (\ref{eq.kms}), 
one gets the expansion
\begin{equation}
\sgh = {(2 \pi)^2 \over Z} \sum_{m,n} (e^{-E_n / T} \pm e^{-E_m / T}) 
|\langle n | J_M(0) | m \rangle|^2 \delta^4(k_\mu - q^n_\mu + q^m_\mu) 
\label{eq.specdef}
\end{equation}
where Z is the partition function, and 
$q^n$ refers to the four-momenta of the state $| n \rangle $.

A stable mesonic state contributes a $\delta$ function-like
peak to the spectral function:
\beq
\sgh = | \langle 0 | J_M | M \rangle |^2 \epsilon(k_0)
\delta(k^2 - m_M^2)
\label{eq.stable}
\eeq
where $m_M$ is the mass of the state. For an unstable
particle one gets a peak with a finite width, e.g., the 
Breit-Wigner form for states with narrow decay width.
We want to study how the spectral function changes due to 
collisions with the thermal medium.

At the level of approximation we are working, the collective excitations,
the fluctuation of the mean field around the vacuum, can be handled within
the Random Phase Approximation (RPA) \cite{fett}. 
In this approximation, which is equivalent to summing over the ring diagrams,
the retarded correlation function is given by 
\begin {equation}
 D_M^R={{\Pi_M} \over{1-2G{\Pi_M}}}.
\label{eq.rpa}
\end {equation}
Here G is the suitable coupling constant and $\Pi_M(k^2)$ is the
one-loop polarisation function for the mesonic channel under consideration,
\begin {equation}
 \Pi_M(k^2) = \int {{d^4p}\over{(2\pi)^4}}{\rm Tr}
              [\Gamma_MS{(p+{k\over 2})}\Gamma_M
              {(p-{k\over 2})}].
\label{eq.pim}
\end {equation}
$S(p)$is the quark propagator and $\Gamma_M$ is the effective
vertex factor. Eqn. (\ref{eq.pim}) has been 
evaluated in the literature; for the NJL model, the 
effective formulae were calculated in ref. \cite{hatsuda}. 
The whole effect of the introduction of the background Polyakov loop 
can be absorbed into a modification of the Fermi-Dirac
distribution functions \cite{hansen}
\begin {eqnarray}
        {f_\Phi^+}(E_p)={ {({\bar\Phi}+2{\Phi}e^{-\beta (E_p+\mu)})}
                          e^{-\beta(E_p+\mu)}+e^{-3\beta (E_p+\mu)}
                     \over 1+3(\bar\Phi+{\Phi} e^{-\beta (E_p+\mu)})
                      e^{-\beta(E_p+\mu)}+e^{-3\beta(E_p+\mu)}}\\
      {f_\Phi^-}(E_p)={ {({\Phi}+2{\bar\Phi}e^{-\beta(E_p-\mu)})}
                          e^{-\beta(E_p-\mu)}+e^{-3\beta(E_p-\mu)}
                     \over 1+3(\Phi+\bar\Phi e^{-\beta(E_p-\mu)})
                      e^{-\beta(E_p-\mu)}+e^{-3\beta(E_p-\mu)}}
\end {eqnarray}
For NJL model,  $\Phi=1$ and the distribution
functions become the usual Fermi-Dirac distribution function.

From the correlation function one can calculate the spectral
function using eq.(\ref{eq.retard}). 
The position of the pole controls the decay of the correlator and is called
the pole mass. It can be obtained by solving 
\begin{equation}
   1-2G{\Pi_M}=0.
\label{eq.pole}
\end{equation}

\begin{table}
\begin{center}
\begin{tabular}{|c|c|}
\hline
$ G_\pi=g_S+g_D \gamma $&$ G_{K^\pm}= g_S + g_D \beta $ \\

$G_{K^0}= g_S + g_D \alpha $&$G^P_{00}=g_S-{2\over 3}{(\alpha+\beta+\gamma)}g_D $ \\

 $G^P_{33}= G_\pi$ & $G^P_{88}=g_S-{1\over 3}{(\gamma-2\alpha
                        -2\beta)g_D}$\\

$G^P_{03}=-{1\over {\sqrt 6}}(\alpha-\beta)g_D$ &
$ G^P_{38}={1\over {\sqrt 3}}(\alpha-\beta)g_D$\\

 $G^P_{08}=-{ {\sqrt 2}\over 6}{(2\gamma-\alpha-\beta)}g_D$ & \\
\hline
\end{tabular}
\caption{Pseudoscalar coupling strengths in $SU(3)$ NJL model,
 where $\alpha=<{\bar u}u>$, $\beta=<{\bar d}d>$, $\gamma=<{\bar s}s>$
\cite{hatsuda,klev}.}
\label{table1}
\end{center}

\end{table}

The coupling constants $G$ for the different flavour combinations are given 
in table 1. The cases of $\eta$ and $\eta^\prime$ need a special 
mention. These mesons arise from the mixing
of the flavour singlet and octet states, {\it i.e.} $\eta_0$
and $\eta_8$ mesons. The mixing arises because of
an interplay between the anomaly term and the nondegeneracy between
$m_s$ and $m_{u,d}$. The masses of $\eta$ and $\eta^\prime$ may 
be obtained from the roots of the equation
\begin{eqnarray}
 \det [{1-2\Pi G}]=\det {\left({{\begin{array}{cc}
  1-(2 \Pi_{00}G_{00}+2\Pi_{80}G_{80})&
  -(2\Pi_{00}{G_{08}}+2\Pi_{08}{G_{88}})\\
   -(2\Pi_{00}{G_{80}}+2\Pi_{88}{G_{80}})&
  1-(2\Pi_{80}{G_{08}}+2\Pi_{88}{G_{88}}) 
\end{array}}}\right)}
           =0 
\label{eq.eta}
\end{eqnarray}
where
\begin {eqnarray}
 \Pi_{00}={2\over 3}[2{ \Pi^P_{uu}}(k)
                      +{ \Pi^P_{ss}}(k)] \nonumber \\
 \Pi_{08}= ({2\sqrt 2\over 3})[{ \Pi^P_{uu}}(k)
                         -{ \Pi^P_{ss}}(k)] \label{eq.etapol} \\
 \Pi_{88}=({2\over 3})[{ \Pi^P_{uu}}(k)
            +2{ \Pi^P_{ss}}(k)] \nonumber
\end {eqnarray}
The polarisation functions are obtained from eq.(\ref{eq.pim}).
On the other hand pole masses of $\eta_0$ and $\eta_8$ can be obtained
directly from $\Pi_{00}$, $\Pi_{88}$ and eqn. (\ref{eq.pole}).
 \vskip 0.3in

\subsection{Results at finite temperatures}
\label{sec.temp}

The pion and sigma channels provide the most interesting observables 
with mesonic states, since they are directly associated with the
chiral symmetry. In the chiral symmetry restored phase, pion and sigma
should be degenerate. Below $\tc$, on the other hand, pion is the
goldstone mode of the spontaneously broken chiral symmetry and is
therefore much lighter than the sigma.

\begin{figure}[htb]
\centerline{\includegraphics[width=5.8in]{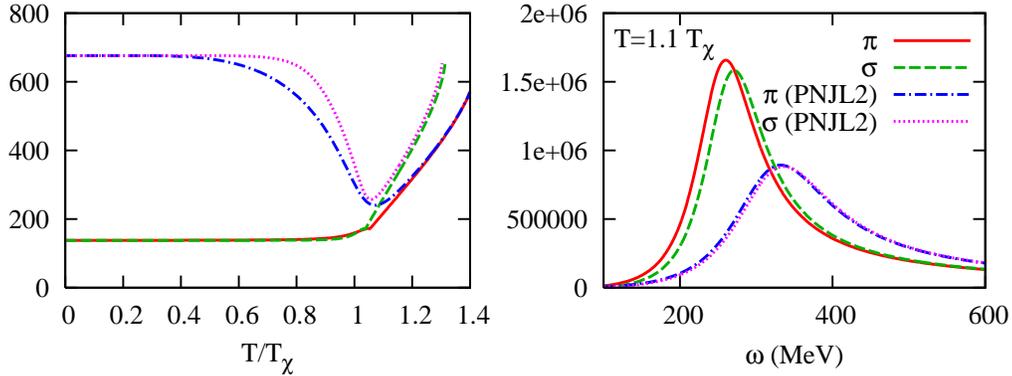}}
\caption{(Color online) (a) Pion and sigma mass versus temperature, 
at $\mu=0$ MeV 
for both NJL (red, solid and blue, dash-dotted) and PNJL (green,dashed 
and pink, dotted)
model. (b) Spectral functions for the pion and sigma in the PNJL model, 
just above $\tc$. Also shown is the effect of suppression of the 
anomaly term at $\tc$ (PNJL2; see Sec. \ref{sec.phase}) on the 
spectral functions .}
\label{fig.pi}
\end{figure}

In fig. \ref{fig.pi} we show the temperature dependence of the pole masses
of the pion and the sigma in the NJL and the PNJL model. We see that close 
to $\tc$ there is substantial difference between the two models.
While the two states come close to each other, they are not degenerate 
just above $\tc$. As mentioned in Sec. \ref{sec.phase},
results of the two models converge at higher temperatures. This,
combined with the fact that $\tc$ is lower for NJL model, leads to a
smoother change in NJL model above $\tc$ than in PNJL. 
The nondegeneracy between the pion and the sigma can also be seen 
in the fig.4(b) , which show the spectral functions in the pion
and sigma channels at a temperature of 1.1 $\tc$. 

There can be two 
sources for the nondegeneracy: the finite (though small) current quark
masses, and the remnant $U_A(1)$ symmetry. To investigate the effect
of the $U_A(1)$ symmetry, we look at PNJL2, which introduces a cutoff 
in $g_D$, as explained in sec. \ref{sec.phase}. As can be seen in
fig. \ref{fig.pi}, this significantly alters the spectral functions,
and removes most of the nondegeneracy of the $\pi$ and $\sigma$ above
$\tc$. Also, while there is a clear structure in both of these
channels above $\tc$, the structure is considerably broader in the
absence of the anomaly term. When studying remnant bound state
structures from QCD-inspired models, it therefore is important to 
keep in mind the effect of the anomaly term, since unlike in QCD, the
anomaly terms are put in here by hand and therefore their
contributions across phase transition may be different from that in QCD. 

\begin{figure}[htb]
\centerline{\includegraphics[width=6in]{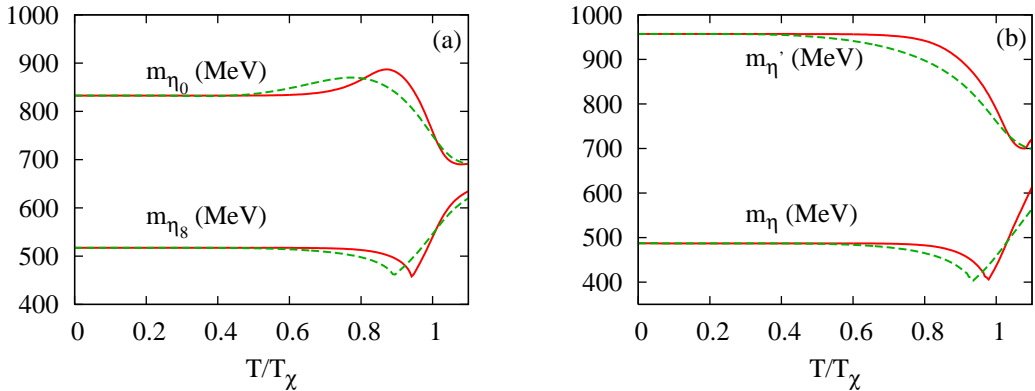}}
\caption{(Color online) Masses of the different $\eta$ states in 
NJL(green, dashed) and PNJL (red, solid) model.}
\label{fig.eta} \end{figure}

In figure \ref{fig.eta} we have plotted the masses of $\eta$, 
$\eta^\prime$, $\eta^0$ and $\eta^8$ mesons respectively. We see features 
similar to those in fig. \ref{fig.pi}: significant differences around
the transition region due to the introduction of the Polyakov loop.
It is also seen that very close to the transition region, the masses
of the singlet and octet states start approaching each other,
indicating that the effect of the anomaly term starts getting reduced.
However, without an explicit cutoff like PNJL2, the nondegeneracy between
these two states is clearly visible at 1.1 $\tc$. 
 
\begin{figure}[htb]
\centerline{\includegraphics[width=6in]{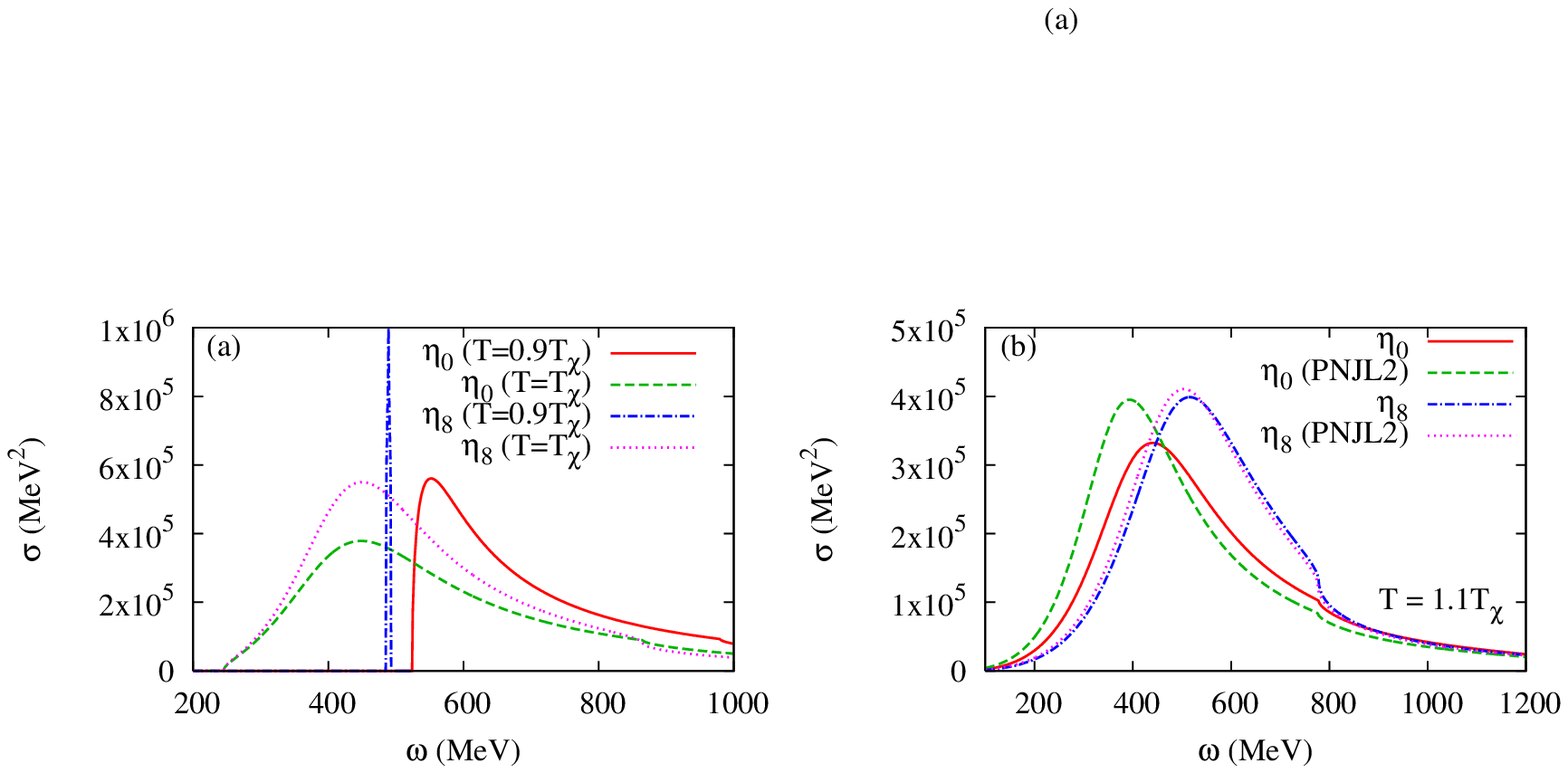}}
\caption{(Color online) (a) Spectral functions of $\eta_{0,8}$ at 
temperatures of 0.9 
$\tc$ and $\tc$. Note the $\delta$ function peak for $\eta_8$ below 
$\tc$. (b) The same above $\tc$. Also shown are the effects of
cutting off the anomaly term above $\tc$ at $\tc$ (PNJL2; see 
Sec. \ref{sec.phase}).}
\label{fig.etaspec}
\end{figure}

The spectral functions of the $\eta_0$ and $\eta_8$ states, calculated
in the PNJL model, are shown in fig. \ref{fig.etaspec}. Here we also 
investigate the effect of the anomaly term by comparing the calculations in 
PNJL2 model.
We find here that the effect of the anomaly
term is considerable in $\eta_0$ but not in $\eta_8$. This can be
understood by looking at the two relevant couplings, $G_{88}$ and $G_{00}$,
in table 1. Because of the different numerical factors, and also the 
different combinations of the quark condensates, the 
anomalous coupling of $\eta_0$ is substantially larger than that of $\eta_8$.
Interestingly, the broadening due to anomaly suppression is
much less even for $\eta_0$ than was seen for the pion. In these
channels, a clear (though broad) resonance structure is seen just
above $\tc$. 

\begin{figure}[htb]
\centerline{\includegraphics[width=3in,height=2in]{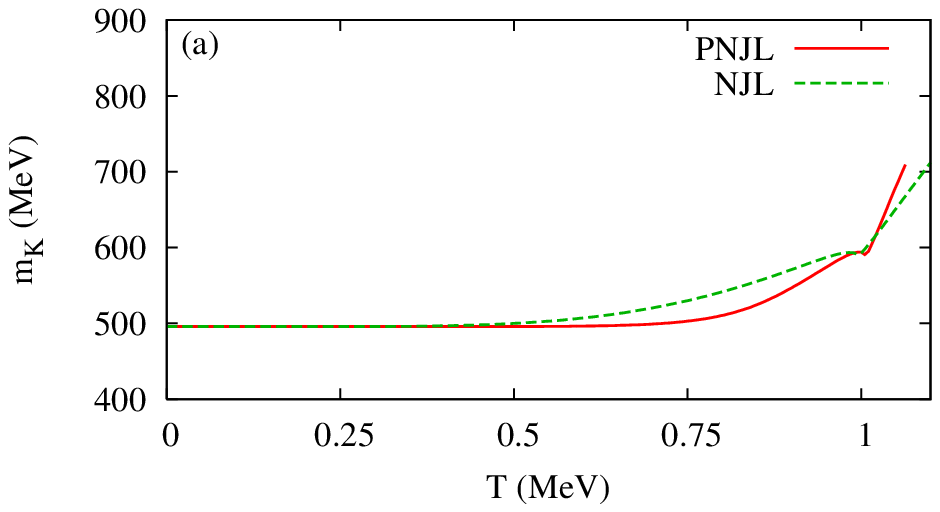}
\includegraphics[width=3in,height=2in]{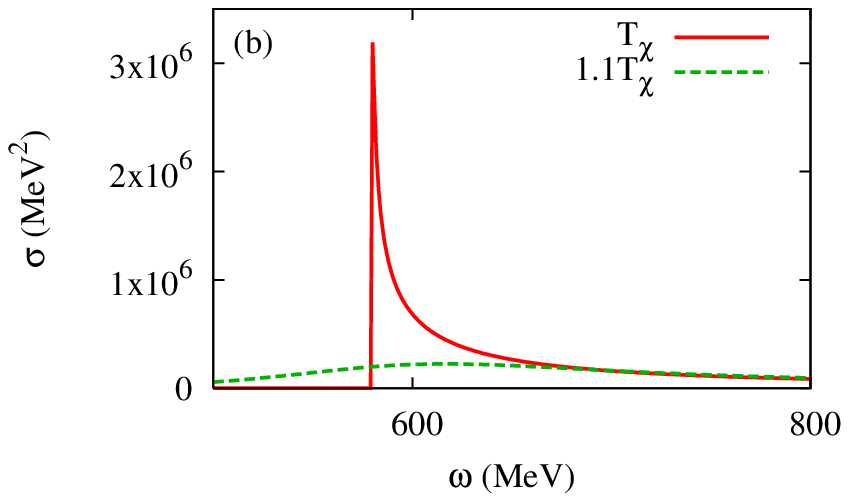}}
\caption{(Color online) (a) Pole mass of the kaon state, in PNJL (solid, red)
and NJL (dashed, green) models, at different temperatures. 
(b) Spectral function, from the PNJL model. The sharp peak at $\tc$
seems to be completely washed off already by 1.1 $\tc$.} 
\label{fig.kaon}
\end{figure}

The behavior of the pole mass for the kaon state, fig. \ref{fig.kaon}, is 
qualitatively similar to the pion. However, the spectral function above $\tc$,
shown in the same figure, shows a very different behavior. The sharp 
peak at $\tc$ is completely washed off already by 1.1 $\tc$: there is no 
indication of a strong correlation between quarks in this channel. 

 {\subsection {Results at finite densities }}
\label{sec.density}

One of the advantages of model studies is that one can study regions
in parameter space which are not easy to study directly from QCD 
in a controlled way. The agreement of PNJL model results with QCD at finite
temperatures encourages us to investigate effects of nonzero baryon
densities, by introducing a quark chemical potential. As mentioned in
Sec. \ref{sec.phase}, however, we do not expect the PNJL model to be a
good model for QCD transition at large $\mq$. We therefore investigate
the mesonic excitations in this model at $\mq$ = 150 MeV. This is
around the region where lattice studies suggest the beginning of the
first order line, while the PNJL model still finds a crossover
transition at these densities. From fig. \ref{fig.phase} we also get 
$\tc \sim$ 170 MeV at $\mq$ = 150 MeV, which is not too far from what
one gets from lattice. 

\begin{figure}[htb]
\centerline{{\hspace{-0.5in}}\includegraphics[width=4in]{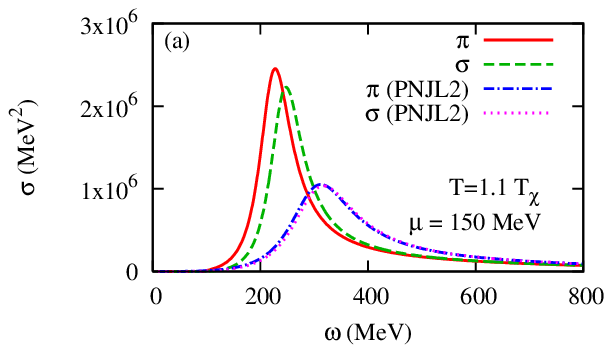}}
\centerline{\includegraphics[width=5in]{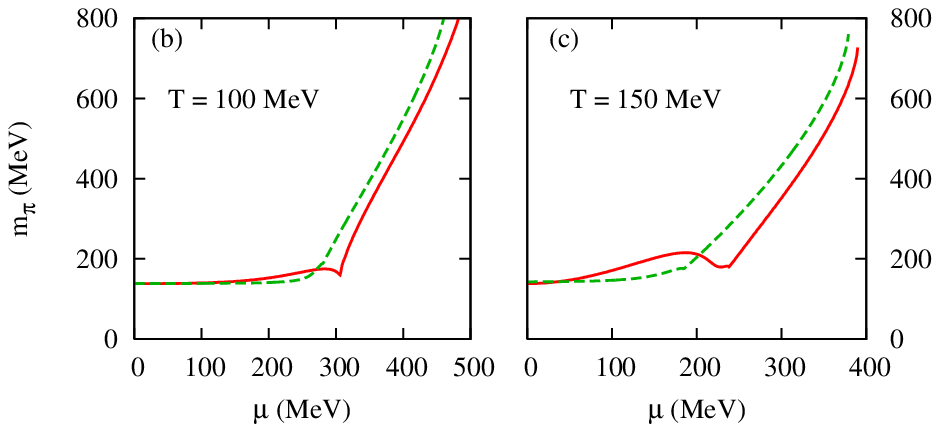}}
\caption{(Color online) (a) Spectral functions of pion near the chiral 
transition point, for $\mq$ = 150 MeV. Also shown is the effect of 
suppression of anomaly term at $\tc$ (PNJL2; see Sec. \ref{sec.phase}).
(b) Pion mass versus $\mu$, 
at $T=100 $MeV. Solid red line is 
for PNJL model and dashed green line is for NJL model. (c) Same as (b) 
but at $T=150$ MeV.}
\label{fig.piden}
\end{figure}

In fig. \ref{fig.piden}, we show the pion and sigma spectral functions above
$\tc$ for $\mq$ = 150 MeV. We see that the spectral
functions have a sharp structure even substantially above $\tc$, as in
the case for $\mq$ = 0. The nondegeneracy of the two channels and the
effect of the anomaly terms are also similar to the $\mq$ = 0 case. 
In the figures 8(b) and 8(c) we show the 
variation of the pion pole mass with $\mq$, for two different
temperatures. Interestingly, the quantitative difference in the masses 
calculated in PNJL and NJL models increases with density. Of course,
as we have discussed above, we do not expect these models to be a good 
approximation to QCD at values of $\mq \gg$ 150 MeV.

\begin{figure}[htb]
\centerline{\includegraphics[width=3.0in]{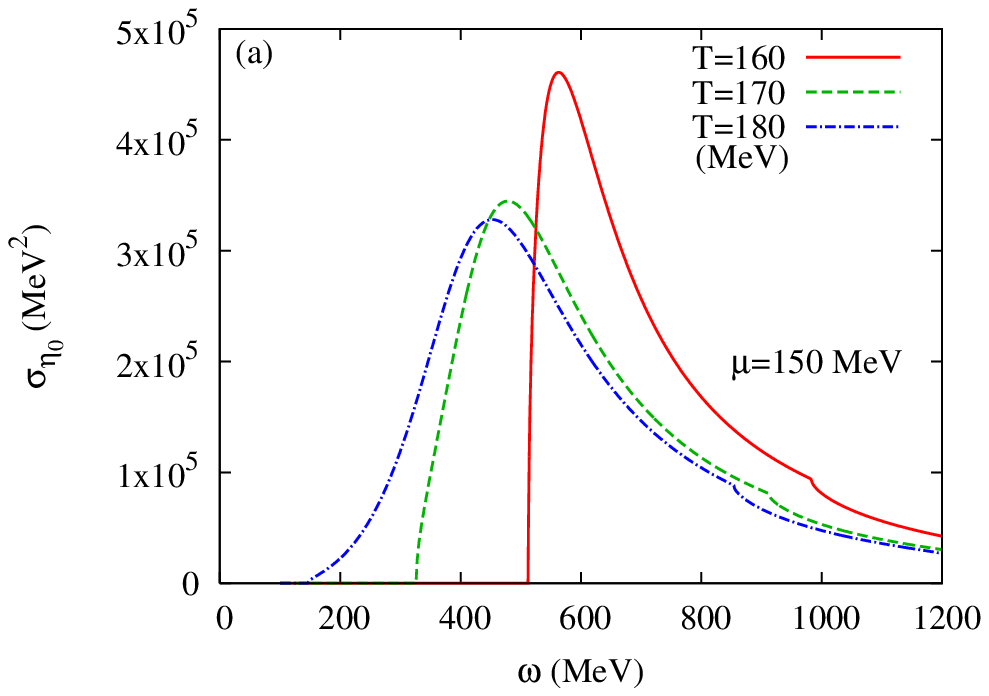}
\includegraphics[width=3.0in]{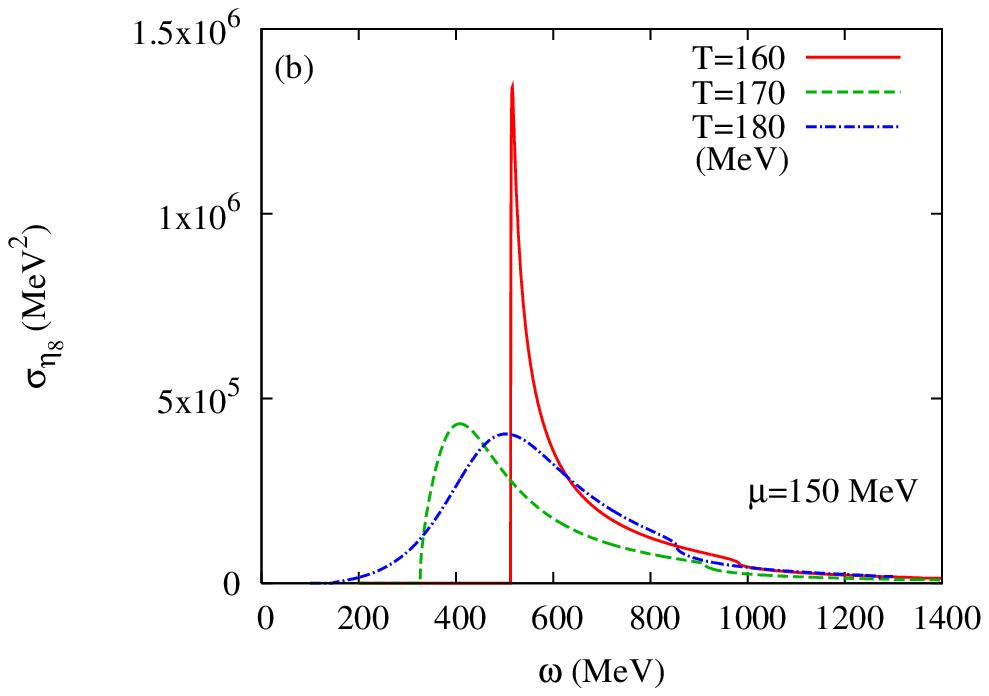}}
\caption{(Color online) (a) The spectral functions of the $\eta_0$ 
   channel around the chiral transition temperatures,
  for $\mq$ = 150 MeV. (b) Same as (a) but for $\eta_8$.}
\label{fig.etaden} \end{figure}

The spectral functions of $\eta_0$ and $\eta_8$ are shown in
fig. \ref{fig.etaden}, for $\mq$ = 150 MeV. This figure has some
interesting features. First, the interplay of the temperature and 
chemical potential gives rise to a nonmonotonous behavior in the peak
of the $\eta_8$ spectral function in the deconfined phase. The peak
position drops sharply at $\tc$, and then rises again. Second, we 
note that the $\eta_8$ peak, which is below the $\eta_0$ peak below
or at $\tc$, becomes very broad already at 1.05 $\tc$, with the
peak position slightly above that of the $\eta_0$ peak. If this
feature is true for full QCD and is not an artefact of the model, it
will have interesting phenomenological implications.

Let us now discuss the sensitivity of our results on enforcement of 
colour neutrality (Sec. \ref{sec.phase}). Our calculations at finite density have 
been carried out at $\mu = 150$ MeV and $T = 1.1 T_c$. At such T and $\mu$,
if we want to enforce color neutrality by the introduction of a color 
chemical potential, we need a color chemical potential $\mu_3 T_3 + \mu_8 T_8$
(in the notation of Ref. \cite{abuki}) with a small $\mu_8 (\sim 30 MeV)$, 
while $\mu_3$ is negligible. Here $T_3$ and $T_8$ are Gell-mann matrices 
in color space. While this changes the individual colour densities, the 
total quark number density ($n_q - n_{\bar q}$) 
and the total scalar density ($n_q + n_{\bar q}$) do not have any appreciable 
change. Similar results about the total quark density has also been reported in 
ref.\cite{abuki1}. In fact the change in the scalar density is 
less than $5\%$ under such conditions. Here we have studied spectral functions 
of colorless, meson-like objects, which are sensitive to the scalar density. 
Hence we do not expect our results at finite chemical potential to be affected 
appreciably by the issue of colour neutrality.

\vskip 0.3in
 {\section {Conclusion}}

To conclude, we have studied the meson-like excitations in the deconfined
plasma at both zero and nonzero quark chemical potential, using the Polyakov 
loop extended Nambu -- Jona-Lasinio (PNJL) model.

The phase diagram of the PNJL model shows a first order line of transitions
at large $\mq$, ending in a critical point at $(\mu_c, T_c) \sim$ (314 MeV, 
92 MeV). We note, however, that the phase diagram can be quite sensitive 
to the temperature dependence of the coefficient of the anomaly term. In 
particular, a suppression of the anomaly term at large temperatures may lead
to a complete wash off of the first order line.   

The in-medium behavior of the mesonic correlations was studied by looking at 
the spectral functions. In the pion and sigma channels, reasonably sharp 
structure was found still at 1.1 $\tc$; also the two states are not degenerate 
at these temperatures. While both these features can be phenomenologically 
interesting, it was also found that these features are quite sensitive to the
anomaly term. In particular, suppressing the anomaly term above $\tc$ leads to
much broader and degenerate structures in these two channels. Since the 
structure of the $\pi$ resonance above $\tc$ is of importance for RHIC 
phenomenology, this sensitivity implores one to look for the restoration 
or otherwise of the $U(A)_1$ anomaly at $\tc$ by looking at suitable 
observables directly on lattice. 

The $\eta_8$ and the $\eta_0$, on the other hand, show much less sensitivity
to the anomaly term. Both these channels show a clear resonance at 1.1 $\tc$.
$\eta_8$ shows very little sensitivity to a suppression of the anomaly term. 
$\eta_0$ is more sensitive, but the resonance-like structure remains 
even when the anomaly term is suppressed. 
In the PNJL model, the kaon channel does not show any sharp structure 
above $\tc$. 
All these behaviors will be of interest for RHIC phenomenology.

We also looked at the mesonic spectral functions just above $\tc$ for a 
nonzero $\mq$. At $\mq$ = 150MeV, which should be in the regime covered by the
RHIC energy scan, PNJL shows a crossover transition. The features of the 
spectral functions in the pion and sigma channels are found to be quite 
similar to those at $\mq$ = 0. A more interesting behavior was found in the 
$\eta$ channels. The interplay of the finite $\mq$ and finite $T$ lead to a 
nonmonotonous behavior of the peak in the $\eta_8$ channel. Also it was found 
that the relative positions of the $\eta_0$ and $\eta_8$ peaks get reversed
above $\tc$: the $\eta_0$ peak comes at a lower $\omega$ than the $\eta_8$
peak at 1.05 $\tc$. It will be interesting to explore the phenomenological 
implications of these behaviours. 

\vskip 0.3in
\acknowledgments{ P.D. thanks CSIR for financial support. A.B. thanks
CSIR for support through the project 03(1074)/06/EMR-II and acknowledges 
UGC UPE grant (computational support). SKG thanks DST, 
Govt. of India for financial support under IRHPA scheme. }


\begin{thebibliography}{99}
\bibitem{note} The effect of the anomaly term on the phase diagram has
also been explored by Fukushima, ref. \cite{fukushima2}. He used different 
values of $g_D$ in eqn (2), and found that for small $g_D$ the first order 
line disappears. In our analysis, the vacuum $g_D$ is set from $\eta^\prime$ 
mass and is not changed; but at finite temperature, we take $g_D(T) = 
g_D {\rm exp}(-T/\tc)^2$.
\bibitem{note2} This is essentially a smoothened version of a step jump
at $\tc$. The factor 5 is chosen arbitrarily so that $g_D(T)/g_D(0)$ is 
similar to PNJL1 at 1.05 $\tc$. Of course, a larger factor would have given 
a sharper drop. Note also that the $U_A(1)$ cutoff, in both PNJL1 and PNJL2, 
is a cutoff at both finite temperature and density, through the $\mu$ dependence
of $\tc$.
\end{thebibliography}
\end{document}